\documentstyle[preprint,aps]{revtex}

\begin{document}
\title{{\bf On the }$B_{c}$ {\bf leptonic decay constant}}
\author{Sameer M. Ikhdair\thanks{%
Electronic address (internet): sameer@neu.edu.tr}}
\address{$^{\ast }$Department of Electrical and Electronic Engineering, Near East\\
University, Nicosia, North Cyprus, Mersin-10, Turkey\\
PACS NUMBER(S): 12.39.Pn, 12.39.Jh, 13.25.Jx, 14.40.-n}
\date{\today
}
\maketitle

\begin{abstract}
We give a review and present a comprehensive calculations for the leptonic
constant $f_{B_{c}}$of the low-lying pseudoscalar and vector states of $%
B_{c} $-meson in the framework of static and QCD-motivated nonrelativistic
potential models taking into account the one-loop and two-loop QCD
corrections in the short distance coefficient that governs the leptonic
constant of $B_{c}$ quarkonium system. Further, we use the scaling relation
to predict the leptonic constant of the ${\rm nS}$-states of the $\overline{b%
}c$ system. Our results are compared with other models to gauge the
reliability of the predictions and point out differences.
\end{abstract}

\bigskip

%
%
%

\section{ INTRODUCTION}

\noindent The $B_{c}$ mesons provide a unique window into the heavy quark
dynamics. Although they are intermediate to the charmonium and bottomonium
systems the properties of $B_{c}$ mesons are a special case in quarkonium
spectroscopy as they are the only quarkonia consisting of heavy quarks with
different flavours. Because they carry flavour they cannot annihilate into
gluons so are more stable and excited $B_{c}$-states lying below ${\rm BD}$
(and ${\rm BD}^{{\rm \ast }}$ or ${\rm B}^{{\rm \ast }}{\rm D}$) threshold
can only undergo radiative or hadronic transitions to the ground state $%
B_{c} $ which then decays weakly. There are two sets of ${\rm S}$-wave
states, as many as two ${\rm P}$-wave multiplets (the ${\rm 1P}$ and some or
all of the ${\rm 2P}$) and one ${\rm D}$-wave multiplet lying below ${\rm BD}
$ threshold for emission of ${\rm B}$ and ${\rm D}$ mesons. As well, the $%
{\rm F}$-wave multiplet is sufficiently close to threshold that they may
also be relatively narrow due to angular momentum barrier suppression of the
Zweig allowed strong decays. However, the spectrum and properties of these
states have been calculated various times in the framework of heavy
quarkonium theory [1].

The discovery of the $B_{c}$ meson by the Collider Detector at Fermilab
(CDF) Collaboration [2] in the channel B$_{c}\longrightarrow J/\psi l\nu $ $%
(l=e,\mu ,\tau )$ with low-lying level of pseudoscalar mass $%
M_{B_{c}}=6.4_{\pm 0.13}^{\pm 0.39}~{\rm GeV}$ and lifetime $\tau
_{B_{c}}=0.46_{-0.16}^{+0.18}\pm 0.03~{\rm ps}$ confirmed the theoretical
predictions regarding various $B_{c}$ meson properties, spectroscopy,
production and decay channels [1].

The description of various aspects of $B_{c}$ meson physics [3] has received
recently a lot of attention both from theoretical as well as experimental
sides. It could offer a unique probe to check the perturbative QCD
predictions more precisely and lead to a new information about the
confinement scale inside hadrons. The higher $\overline{b}c$-meson state
cascades down through lower energy $\overline{b}c$ state via hadronic or
electromagnetic transitions to the pseudoscalar ground state $B_{c}$ meson.
The theoretical study of the pure leptonic decays of the $B_{c}$ meson such
as $B_{c}\longrightarrow l\nu _{l\text{ }}$can be used to determine the
leptonic decay constant$\ f_{B_{c}}$ [4], as well as the fundamental
parameters in the standard Model (SM). Hence, it is the only meson within SM
which is composed of two nonrelativistic heavy quarks of different flavors:
open charm and bottom quarks. Its spectroscopy production mechanisms and
decays differ significantly from those of charmonium $J/\psi $ and upsilon $%
\Upsilon $ as well as hadrons with one heavy quark. Indeed, this meson is
also a long lived system decaying through electroweak interactions. It
stands among the families of $\ \overline{c}c$ and $\overline{b}b$ and thus
could be used to study both quantitatively and conceptually existing
effective low energy frameworks for the description of bound state heavy
quark dynamics, like NRQCD [5], pNRQCD [6] and vNRQCD [7].

The calculation of the $B_{c}$ leptonic decay constant $\ f_{B_{c}}$ can be
carried out either using QCD-based methods, such as lattice QCD [8], QCD sum
rules [9], or adopting some constituent quark models [1]. So far, lattice
QCD has only been employed to calculate the $B_{c}$ purely leptonic width.
As the QCD sum rules, the $B_{c}$ leptonic constant, as well as the matrix
elements relevant for the semileptonic decays were computed. Moreover, the
leptonic constant was also been calculated by using the nonrelativistic (NR)
potential model [1].

Ikhdair {\it et al}. [10] applied the NR form of the statistical model to
calculate the spectroscopy, decay constant and some other properties of the
heavy mesons, including the $\overline{b}c$ system, using a class of three
static quarkonium potentials. Davies {\it et al.} [8] predicted the leptonic
constant of the lowest state of the $B_{c}$ system with the recent lattice
calculations. Eichten and Quigg [1] gave a more comprehensive account of the
decays of the $B_{c}$ system that was based on the QCD-motivated potential
of Buchm\"{u}ller and Tye [11]. Gershtein {\it et al. }[1] also published a
detailed account of the energies and decays of the $B_{c}$ system using a
QCD sum-rule calculations [9]. Fulcher [2] also calculated the leptonic
constant of the $B_{c}$ quarkonium system using the treatment of the
spin-dependent potentials to the full radiative one-loop level and thus
included effects of the running coupling constant in these potentials.
Furthermore, he also used the renormalization scheme developed by Gupta and
Radford [12]. Ebert {\it et al}. [1] have also calculated the leptonic
constant of the pseudoscalar and vector $B_{c}$ quarkonium system using a
relativistic model and then compared their calculation with the NR model.
Kiselev {\it et al}. [13-15] calculated the leptonic constant\ of the
pseudoscalar meson $B_{c}$ in the framework of QCD-motivated potential
models taking into account the one-loop [13], the two-loop [14] correction
matching and scaling relation (SR) [15]. Capstick and Godfrey [16] predicted
the result of $f_{B_{c}}=410$ $MeV$ for the decay constant of the $B_{c}$
meson using the Mock-Meson approach or other relativistic quark models.
Ikhdair and Sever [17] have calculated the decay constant in the
nonrelativistic and semi-relativistic quark model using the shifted large-$%
{\rm N}$ expansion technique. Finally, Godfrey [18] has calculated the
spectroscopy of $B_{c}$ meson in the relativixzed quark model. Motyka and
Zalewski [18] proposed a nonrelativistic Hamiltonian with plausible spin
dependent corrections for the quarkonia below their respective strong decay
threshold. They found the decay constants for the ground state and for the
first excited ${\rm S}$-state of $\overline{b}c$ quarkonium to be $%
f_{B_{c}}=435$ $MeV$ and $f_{B_{c}}=315$ $MeV,$ respectively.$.$

We have made this study mostly a fully treatment for the quarkonium
potential models used in the literature [1,17] in calculating the $B_{c}$
leptonic constant based on the NR potential model, (cf. e.g. [17] and
references contained there), and applied with a great success to fit the
entire heavy quarkonum spectroscopy [17,19]. We insist upon strict
flavor-independence of its parameters. We also use the potential models to
give a simultaneous account of the properties of the $\overline{c}c$, $%
\overline{b}b$ and $\overline{b}c$ quarkonium systems.

The contents of this paper is as follows: In section II, we present briefly
the solution of the Schr\"{o}dinger equation for the $\overline{b}c$
quarkonium mass spectrum using the shifted large-N-expansion technique
(SLNET). In section III, we introduce the necessary expressions for the
one-loop and two-loop corrections to the leptonic constant available to
moment. Further, in section IV we present the phenomenological and the
QCD-motivated potentials used in the present work. Finally, section V
contains our calculations and remarks made for the leptonic constant in our
approach and scaling relation.

\section{WAVE EQUATION}

\noindent In this section we consider the ${\rm N}$-dimensional space
Schr\"{o}dinger equation for any spherically symmetric potential $V({\bf r})$%
. If $\psi ({\bf r})$ denotes the Schr\"{o}dinger's wave function, a
separation of variables $\psi ({\bf r})=Y_{l,m}(\theta ,\phi
)u(r)/r^{(N-1)/2}$ gives the following radial equation (in units $\hbar
=c=1) $ [17,19,20]

\begin{equation}
\left\{ -\frac{1}{4\mu }\frac{d^{2}}{dr^{2}}+\frac{[\overline{k}-(1-a)][%
\overline{k}-(3-a)]}{16\mu r^{2}}+V(r)\right\} u(r)=E_{n,l}u(r),  \label{1}
\end{equation}
where $\mu =\left( m_{\overline{q}_{1}}m_{q_{2}}\right) /(m_{\overline{q}%
_{1}}+m_{q_{2}})$ is the reduced mass for the two interacting particles.
Here $E_{n,l}$ denotes the Schr\"{o}dinger binding energy and $\overline{k}%
=N+2l-a,$ with $a$ representing a proper shift to be calculated later on and 
$l$ is the angular quantum number. We follow the shifted $1/N$ or $1/%
\overline{k}$ expansion method [17] by defining

\begin{equation}
V(y(r_{0}))\;=\frac{1}{Q}\stackrel{\infty }{%
\mathrel{\mathop{\sum }\limits_{m=0}}%
}\left( \frac{d^{m}V(r_{0})}{dr_{0}^{m}}\right) \frac{\left( r_{0}y\right)
^{m}}{m!}\overline{k}^{-(m-4)/2},  \label{2}
\end{equation}
and also the energy eigenvalue expansion [17,19]

\begin{equation}
E_{n,l}\;=\stackrel{\infty }{%
\mathrel{\mathop{\sum }\limits_{m=0}}%
}\frac{\overline{k}^{(2-m)}}{Q}E_{m},  \label{3}
\end{equation}
where $x=\overline{k}^{1/2}(r/r_{0}-1)$ with $r_{0}$ is an arbitrary point
where the Taylor expansion is being performed about and $Q$ is a scale to be
set equal to $\overline{k}^{2}$ at the end of our calculations. Inserting
equations (2) and (3) into equation (1) gives

\[
\left[ -\frac{1}{4\mu }\frac{d^{2}}{dy^{2}}+\frac{1}{4\mu }\left( \frac{\bar{%
k}}{4}-\frac{(2-a)}{2}+\frac{(1-a)(3-a)}{4\bar{k}}\right) \times \stackrel{%
\infty }{%
\mathrel{\mathop{\sum }\limits_{m=0}}%
}(-1)^{m}\frac{\left( 1+m\right) }{\overline{k}^{m/2}}y^{m}\right. 
\]

\begin{equation}
+\frac{r_{0}^{2}}{Q}\left. \stackrel{\infty }{%
\mathrel{\mathop{\sum }\limits_{m=0}}%
}\left( \frac{d^{m}V(r_{0})}{dr_{0}^{m}}\right) \frac{\left( r_{0}y\right)
^{m}}{m!}\overline{k}^{-(m-2)/2}\right] \chi _{n_{r}}(y)=\xi _{n_{r}}\chi
_{n_{r}}(y),  \label{4}
\end{equation}
where the final analytic expression for the $1/\overline{k}$ expansion of
the energy eigenvalues appropriate to the Schr\"{o}dinger particle is [17] 
\begin{equation}
\xi _{n_{r}}=\frac{r_{0}^{2}}{Q}\stackrel{\infty }{%
\mathrel{\mathop{\sum }\limits_{m=0}}%
}\overline{k}^{(1-m)}E_{m},  \label{5}
\end{equation}
where $n_{r}$ is the radial quantum number. Hence, we formulate the SLNET
(expansion as $1/\bar{k})$ for the nonrelativistic motion of spinless
particle bound in spherically symmetric potential V(r). The resulting
eigenvalue of the N-dimensional Schr\"{o}dinger equation is written as [17]

\begin{eqnarray}
\xi _{n_{r}} &=&\bar{k}\left[ \frac{1}{16\mu }+\frac{r_{0}^{2}V(r_{0})}{Q}%
\right] +\left[ (n_{r}+\frac{1}{2})\;\omega -\frac{(2-a)}{8\mu }\right] 
\nonumber \\
&&+\frac{1}{\overline{k}}\left[ \frac{(1-a)(3-a)}{16\mu }+\alpha ^{(1)}%
\right] +\frac{\alpha ^{(2)}}{\overline{k}^{2}},  \label{6}
\end{eqnarray}
where $\alpha ^{(1)}$ and $\alpha ^{(2)}$ are two useful expressions given
by Imbo et al [20]. Comparing equation (5) with (6) yields

\begin{equation}
E_{0}=V(r_{0})+\frac{Q}{16\mu r_{0}^{2}},  \label{7}
\end{equation}
\begin{equation}
E_{1}=\frac{Q}{r_{0}^{2}}\left[ \left( n_{r}+\frac{1}{2}\right) \omega -%
\frac{(2-a)}{8\mu }\right] ,  \label{8}
\end{equation}
\begin{equation}
E_{2}=\frac{Q}{r_{0}^{2}}\left[ \frac{(1-a)(3-a)}{16\mu }+\alpha ^{(1)}%
\right] ,  \label{9}
\end{equation}
and 
\begin{equation}
E_{3}=\frac{Q}{r_{0}^{2}}\alpha ^{(2)}.  \label{10}
\end{equation}
The quantity $r_{0}$ is chosen so as to minimize the leading term, $E_{0}$
[17,19], that is,

\begin{equation}
\frac{dE_{0}}{dr_{0}}=0\text{ \ \ \ and \ \ }\frac{d^{2}E_{0}}{dr_{0}^{2}}>0.
\label{11}
\end{equation}
Therefore, $r_{0}$ satisfies the relation

\begin{equation}
Q=8\mu r_{0}^{3}V^{\prime }(r_{0}),  \label{12}
\end{equation}
and to solve for the shifting parameter $a$, the next contribution to the
energy eigenvalue $E_{1}$ is chosen to vanish [20] so that the second- and
third-order corrections are very small compared with the leading term
contribution. The energy states are calculated by considering the leading
term $E_{0\text{ }}$and the second-order and third-order corrections, it
implies the shifting parameter 
\begin{equation}
a=2-(2n_{r}+1)\left[ 3+\frac{r_{0}V^{\prime \prime }(r_{0})}{V^{\prime
}(r_{0})}\right] ^{1/2}.  \label{13}
\end{equation}
Therefore, the Schr\"{o}dinger binding energy to the third order is

\begin{equation}
E_{n,l}=V(r_{0})+\frac{r_{0}V^{\prime }(r_{0})}{2}+\frac{1}{r_{0}^{2}}\left[ 
\frac{(1-a)(3-a)}{16\mu }+\alpha ^{(1)}+\frac{\alpha ^{(2)}}{\overline{k}}%
+O\left( \frac{1}{\overline{k}^{2}}\right) \right] ,  \label{14}
\end{equation}
Once the problem is collapsed to its actual dimension ($N=3),$ it simply
rests the task of relating the coefficients of our equation to the
one-dimensional anharmonic oscillator in order to read the energy spectrum.
For the $N=3$ physical space, the equation (1) restores its
three-dimensional form. Thus, with the choice $\overline{k}=\sqrt{Q}$ which
rescales the potential, we derive an analytic expression that satisfies $%
r_{0}$ in equations (12)-(13) as

\begin{equation}
1+2l+(2n_{r}+1)\left[ 3+\frac{r_{0}V^{\prime \prime }(r_{0})}{V^{\prime
}(r_{0})}\right] ^{1/2}=\left[ 8\mu r_{0}^{3}V^{\prime }(r_{0})\right]
^{1/2}.  \label{15}
\end{equation}
where $n_{r}=0,1,2,\cdots $ stands for the radial quantum number and $%
l=0,1,2,\cdots $ stands for the angular quantum number. Once $r_{0}$ is
being determined through equation (15), the Schr\"{o}dinger binding energy
of the $\overline{q}_{1}q_{2}$ system in equation (14) becomes relatively
simple and straightforward. We finally obtain the total Schr\"{o}dinger mass
binding energy for spinless particles as

\begin{equation}
M(\overline{q}_{1}q_{2})=m_{\overline{q}_{1}}+m_{q_{2}}+2E_{n,l}.  \label{16}
\end{equation}
As stated before in [17,19,20], for any fixed $n$ the computed energies
become more convergent as $l$ increases. This is expected since the
expansion parameter $1/N$ or $1/\overline{k}$ becomes smaller as $l$ becomes
larger since the parameter $\overline{k}$ is proportional to $n$ and appears
in the denominator in higher-order correction.

On the other hand, the spin dependent correction to the nonrelativistic
Hamiltonian, which is responsible for the hyperfine splitting of the ${\rm 1S%
}$-state mass level is generally used in the form (cf. e.g. Refs. [1,17,21])%
\footnote{%
To the moment, the only measured splitting of $nS$-levels is that of $\eta
_{c}$ and $J/\psi ,$ which allows us to evaluate the so-called SAD using $%
\overline{M}_{\psi }({\rm 1S})=(3M_{J/\psi }+M\eta _{c})/4$ and also $%
\overline{M}({\rm nS})=M_{V}({\rm nS})-(M_{J/\psi }-M\eta _{c})/4n$ [13,21].}

\begin{equation}
\Delta E_{{\rm HF}}=\frac{32\pi \alpha _{s}(\mu )}{9m_{c}m_{b}}\left| \psi _{%
{\rm 1S}}(0)\right| ^{2}.  \label{17}
\end{equation}
Like most authors (cf. e.g. [1,17,21]), we determine the coupling constant $%
\alpha _{s}(m_{c})$ from the well measured experimental hyperfine splitting
of the ${\rm 1S}(c\overline{c})$ state value 
\begin{equation}
\Delta E_{{\rm HF}}=M_{J/\psi }-M_{\eta _{c}}=117\pm 2~{\rm MeV.}  \label{18}
\end{equation}
We use the value of the coupling constant to reproduce the spin-averaged
data (SAD) or center-of-gravity (c.o.g.) of the lowest charmonium state
value, that is, $\overline{M}_{\psi }({\rm 1S}).$ In order to apply this
formula one needs the value of the wave function at the origin, this is
obtained by solving the Schr\"{o}dinger equation with the nonrelativistic
Hamiltonian and the coupling constant $\alpha _{s}(\mu ).$ The radial wave
function at the origin [21] is determined by

\begin{equation}
\left| R_{{\rm 1S}}(0)\right| ^{2}=2\mu \left\langle \frac{dV(r)}{dr}%
\right\rangle .  \label{19}
\end{equation}
where $\left| R_{{\rm 1S}}(0)\right| ^{2}=4\pi \left| \psi _{{\rm 1S}%
}(0)\right| ^{2}.$ Hence, the total mass of the low-lying pseudoscalar $%
B_{c} $-state is given by the expression

\begin{equation}
M_{B_{c}}(0^{-})=m_{c}+m_{b}+2E_{1,0}-3\Delta E_{{\rm HF}}/4,  \label{20}
\end{equation}
and also for the vector $B_{c}^{\ast }$-state

\begin{equation}
M_{B_{c}^{\ast }}(1^{-})=m_{c}+m_{b}+2E_{1,0}+\Delta E_{{\rm HF}}/4.
\label{21}
\end{equation}
Finally, the square-mass difference can be simply found by using the
expression

\begin{equation}
\Delta M^{2}=M_{B_{c}^{\ast }}^{2}(1^{-})-M_{B_{c}}^{2}(0^{-})=2\Delta E_{%
{\rm HF}}\left[ m_{c}+m_{b}+2E_{1,0}-\Delta E_{{\rm HF}}/4\right] .
\label{22}
\end{equation}

\section{Leptonic decay constant of the $B_{c}$-meson}

\noindent The study of the heavy quarkonium system has played a vital role
in the development of the QCD. Some of the earliest applications of
perturbative QCD were calculations of the decay rates of charmonium [22].
These calculations were based on the assumption that, in the NR limit, the
decay rate factors into a short-distance (SD) perturbative part associated
with the annihilation of the heavy quark and antiquark and a long-distance
(LD) part associated with the quarkonium wavefunction. Calculations of the
annihilation decay rates of heavy quarkonium have recently been placed on a
solid theoretical foundation by Bodwin {\it et al}. [23]. Their approach is
based on NRQCD, and effective field theory that is equivalent to QCD to any
given order in the relative velocity $v$ of the heavy quark and antiquark
[24]. Using NRQCD [5] to seperate the SD and LD effects, Bodwin {\it et al}.
derived a general factorization formula for the inclusive annihilation decay
rates of heavy quarkonium. The SD factors in the factorization formula can
be calculated using pQCD, and the LD factors are defined rigorously in terms
of the matrix elements of NRQCD that can be estimated using lattice
calculations. It applies equally well to ${\rm S}$-wave, ${\rm P}$-wave, and
higher orbital-angular-momentum states, and it can be used to incorporate
relativistic corrections to the decay rates.

In the NRQCD [25] approximation for the heavy quarks, the calculation of the
leptonic decay constant for the heavy quarkonium with the two-loop accuracy
requires the matching of NRQCD currents with corresponding full-QCD
axial-vector currents [14]

\begin{equation}
\left. {\cal J}^{\lambda }\right| _{{\rm NRQCD}}=-\chi _{b}^{\dagger }\psi
_{c}v^{\lambda }~\ \text{and \ }\left. J^{\lambda }\right| _{{\rm QCD}}=%
\overline{b}\gamma ^{\lambda }\gamma _{5}c,  \label{23}
\end{equation}
where $b$ and $c$ are the relativistic bottom and charm fields,
respectively, $\chi _{b}^{\dagger }$ and $\psi _{c}$ are the NR spinors of
anti-bottom and charm and $v^{\lambda }$ is the four-velocity of heavy
quarkonium. The \ NRQCD [25] lagrangian describing the $B_{c}$-meson bound
state dynamics is

\begin{equation}
{\cal L}_{{\rm NRQCD}}={\cal L}_{{\rm light}}+\psi _{c}^{\dagger }\left(
iD_{0}+{\bf D}^{2}/(2m_{c})\right) \psi _{c}+\chi _{b}^{\dagger }\left(
iD_{0}-{\bf D}^{2}/(2m_{b})\right) \chi _{b}+\;\cdots ,  \label{24}
\end{equation}
where ${\cal L}_{{\rm light}}$ is the relativistic lagrangian for gluons and
light quarks. The two-component spinor field $\psi _{c}$ annihilates charm
quarks, while $\chi _{b}$ creates bottom anti-quarks. The relative velocity $%
v$ of heavy quarks inside the $B_{c}$-meson provides a small parameter that
can be used as a nonperturbative expansion parameter. To express the decay
constant $f_{B_{c}}$ in terms of NRQCD matix elements we express\ $\left.
J^{\lambda }\right| _{{\rm QCD}}$ in\ terms of NRQCD fields $\psi _{c}$ and $%
\chi _{b}.$ The $\lambda =0$ current-component contributes to the matrix
element and consequently the $\left. J^{\lambda }\right| _{{\rm QCD}}$ has
the following operator expansion

\begin{equation}
\left\langle 0\right| \overline{b}\gamma ^{\lambda }\gamma _{5}c\left| B_{c}(%
{\bf P})\right\rangle =if_{B_{c}}P^{\lambda },  \label{25}
\end{equation}
where $\left| B_{c}({\bf P})\right\rangle $ is the state of the $B_{c}$%
-meson with four-momentum $P.$ Only the $\lambda =0$ component contributes
to the matrix element (25) in the rest frame of the $B_{c}$-meson. It has
the standard covariant normalization

\begin{equation}
\frac{1}{(2\pi )^{3}}\int \psi _{B_{c}}^{\ast }(p^{\prime })\psi
_{B_{c}}^{{}}(p)d^{3}p=2E\delta ^{(3)}(p^{\prime }-p),  \label{26}
\end{equation}
and its phase has been chosen so that $f_{B_{c}}$ is real and positive. In
NRQCD, the matching yields

\begin{equation}
\overline{b}\gamma ^{0}\gamma _{5}c=K_{0}\chi _{b}^{\dagger }\psi _{c}+K_{2}(%
{\bf D}\chi _{b})^{\dagger }.{\bf D}\psi _{c}+\;\cdots ,  \label{27}
\end{equation}
where $K_{0}=K_{0}(m_{c},m_{b}$) and $K_{2}=K_{2}(m_{c},m_{b}$) are Wilson
SD coefficients. They can be determined by matching perturbative
calculations of the matrix element $\left\langle 0\right| \overline{b}\gamma
^{0}\gamma _{5}c\left| B_{c}\right\rangle $ that is mainly resulting from
the operator $\chi _{b}^{\dagger }\psi _{c}$ in 
\begin{equation}
\left. \left\langle 0\right| \overline{b}\gamma ^{0}\gamma _{5}c\left|
B_{c}\right\rangle \right| _{{\rm QCD}}=\left. \left. K_{0}\left\langle
0\right| \chi _{b}^{\dagger }\psi _{c}\left| B_{c}\right\rangle \right| _{%
{\rm NRQCD}}+K_{2}\left\langle 0\right| ({\bf D}\chi _{b})^{\dagger }.{\bf D}%
\psi _{c}\left| B_{c}\right\rangle \right| _{{\rm NRQCD}}+\;\cdots ,
\label{28}
\end{equation}
where the matrix element on the left side of (28) is taken between the
vacuum and the state $\left| B_{c}\right\rangle .$\ Hence, equation (28) can
be estimated as

\begin{equation}
\left| \left\langle 0\right| \chi _{b}^{\dagger }\psi _{c}\left|
B_{c}\right\rangle \right| ^{2}\simeq \frac{3M_{B_{c}}}{\pi }\left|
R_{1S}(0)\right| ^{2}.  \label{29}
\end{equation}
Onishchenko and Veretin [25] calculated the matrix elements on both sides of
\ equation (28) up to $\alpha _{s}^{2}$ order. In one-loop calculation, the
SD-coefficients are 
\begin{equation}
K_{0}=1\text{ \ and \ K}_{2}=-\frac{1}{8\mu ^{2}},  \label{30}
\end{equation}
with $\mu $ defined after equation (1). Further, Braaten and Fleming in
their work [26] calculated the perturbation correction to $K_{0}$ up to
order- $\alpha _{s}$ (one-loop correction) as

\begin{equation}
K_{0}=1+c_{1}\frac{\alpha _{s}(\mu )}{\pi },  \label{31}
\end{equation}
with $c_{1}$ being calculated in Ref. [26] as 
\begin{equation}
c_{1}=-\left[ 2-\frac{m_{b}-m_{c}}{m_{b}+m_{c}}\ln \frac{m_{b}}{m_{c}}\right]
.  \label{32}
\end{equation}
Finally, the leptonic decay constant for the one-loop calculations is 
\begin{equation}
f_{B_{c}}^{{\rm (1-loop)}}=\left[ 1-\frac{\alpha _{s}(\mu )}{\pi }\left[ 2-%
\frac{m_{b}-m_{c}}{m_{b}+m_{c}}\ln \frac{m_{b}}{m_{c}}\right] \right]
f_{B_{c}}^{{\rm NR}},  \label{33}
\end{equation}
where the NR leptonic constant [27] is

\begin{equation}
f_{B_{c}}^{{\rm NR}}=\sqrt{\frac{3}{\pi M_{B_{c}}}}\left| R_{{\rm 1S}%
}(0)\right|  \label{34}
\end{equation}
and $\mu $ is any scale of order $m_{b}$ or $m_{c}$ of the running coupling
constant. On the other hand, the calculations of two-loop correction in the
case of vector current and equal quark masses was done in [28]. Further,
Onishchenko and Veretin [25] extended the work of Ref. [28] into the the
non-equal mass case. They found an expression for the two-loop QCD
corrections to $B_{c}$-meson leptonic constant given by

\begin{equation}
K_{0}(\alpha _{s},M/\mu )=1+c_{1}(M/\mu )\frac{\alpha _{s}(M)}{\pi }%
+c_{2}(M/\mu )\left( \frac{\alpha _{s}(M)}{\pi }\right) ^{2}+\cdots ,
\label{35}
\end{equation}
where $c_{2}(M/\mu )$ is the two-loop matching coefficient and with $c_{1,2}$
are explicitly given in Eq. (32) and (Ref. [25]; see equations (16)-(20)
therein), respectively. In the case of $B_{c}$-meson and pole quark masses $%
(m_{b}=4.8~{\rm GeV},~m_{c}=1.65~{\rm GeV}),$ they found 
\begin{equation}
f_{B_{c}}^{{\rm (2-loop)}}=\left[ 1-1.48\left( \frac{\alpha _{s}(m_{b})}{\pi 
}\right) -24.24\left( \frac{\alpha _{s}(m_{b})}{\pi }\right) ^{2}\right]
f_{B_{c}}^{{\rm NR}}.  \label{36}
\end{equation}
Therefore, the two-loop corrections are large and constitute nearly $100\%$
of one-loop correction as stated in Ref. [25].

\section{SOME POTENTIAL MODELS}

\noindent The potential models are suitable for the phenomenological
studies, because they can reproduce the model formulae or numbers for the
quantities, used as input values (the level masses, for example). Hence the
potential models can be considered as phenomenological meaningful fittings
of some experimental values, but they can not restore a true potential, that
does not exist due to the nonperturbative effects [15].

\subsection{Static potentials}

It is easy to see that most phenomenological static potentials used in the
Scr\"{o}dinger equation in [17] may be gathered up in a general form [17,29] 
\begin{equation}
V(r)=-ar^{-\alpha }+br^{\beta }+V_{0}\text{ \ \ \ \ }0\leq \alpha ,\beta
\leq 1,\text{ \ \ \ \ }a,b\geq 0,\text{ \ }  \label{37}
\end{equation}
where $V_{0}$ may be of either sign. Here we assume that the effective $%
\overline{q}_{1}q_{2}$ potential consists of two terms, one of which, $%
V_{V}(r)=-ar^{-\alpha },$ transforms like a time-component of a Lorentz
4-vector and the other, $V_{S}(r)=br^{\beta },$ like a Lorentz scalar. These
static quarkonium potentials are monotone nondecreasing, and concave
functions which satisfy the condition

\begin{equation}
V^{\prime }(r)>0\text{ \ and }V^{\prime \prime }(r)\leq 0.  \label{38}
\end{equation}
\ \ At least ten potentials of this general form, but with various values of
the parameters, have been proposed in the literature (cf. [17] and
references contained there). The generality (37) comprises the following
five types of potentials used in the literature:\ Some of these potentials
have $\alpha =\beta $ as in the Cornell ($\alpha =\beta =1)$; Song-Lin ($%
\alpha =\beta =\frac{1}{2})$\ and Turin ($\alpha =\beta =\frac{3}{4})$
potentials with same sets of fitting parameters used in our previous works
[17]. Further, Song [30] has also used a potential with $\alpha =\beta =%
\frac{2}{3}.$ \ \ \ \ \ \ \ \ \ \ \ \ \ \ \ \ \ \ \ \ \ \ \ \ \ \ \ \ \ \ \
\ \ \ \ \ \ \ \ \ \ \ \ \ \ \ \ \ \ \ \ \ \ \ \ \ \ \ \ \ \ \ \ \ \ \ \ \ \
\ \ \ \ \ \ \ \ \ \ \ \ \ \ \ \ \ \ \ \ \ \ \ \ \ \ \ \ \ \ \ \ \ \ \ \ \ \
\ \ \ \ \ \ \ \ \ \ \ \ \ \ \ \ \ \ \ \ \ \ \ \ \ \ \ \ \ \ \ \ \ \ \ 

On the other hand, potentials with $\alpha \neq \beta $ have also been
popular as

\subsubsection{\it Martin potential}

The phenomenological power-law potential proposed by Martin (cf. e.g.
[15,17]) has $\alpha =0,$ $\beta =0.1$ of the form

\begin{equation}
V_{{\rm M}}(r)=b_{{\rm M}}(\Lambda _{{\rm M}}r)^{0.1}+c_{{\rm M}},
\label{39}
\end{equation}
is labeled as Martin's potential [15] with a given set of adjustable
parameters 
\begin{equation}
\left[ b_{{\rm M}},c_{{\rm M}},\Lambda _{{\rm M}}\right] =\left[ 6.898\text{ 
}{\rm GeV}^{{\rm 1.1}},-8.093\text{ }{\rm GeV},1~{\rm GeV}\right] ,
\label{40}
\end{equation}
and quark masses

\begin{equation}
\left[ m_{c},m_{b}\right] =\left[ 1.800~{\rm GeV},5.174~{\rm GeV}\right] .
\label{41}
\end{equation}
(potential units are also in ${\rm GeV}$). \ \ \ \ \ \ \ \ \ \ \ \ \ \ \ \ \
\ \ \ \ \ \ \ \ \ \ \ \ \ \ \ \ \ \ \ \ \ \ \ \ \ \ \ \ \ \ \ \ \ \ \ \ \ \
\ \ \ \ \ \ \ \ \ \ 

\subsubsection{\it Logarithmic potential}

A Martin's power-law potential\ turns into the logarithmic potential of
Quigg and Rosner [15,17] corresponds to $\alpha =\beta \rightarrow 0$ and it
takes the general form

\begin{equation}
V_{{\rm L}}(r)=b_{{\rm L}}\ln (\Lambda _{{\rm L}}r)+c_{{\rm L}},  \label{42}
\end{equation}
with an adjustable set of parameters 
\begin{equation}
\left[ b_{{\rm L}},c_{{\rm L}},\Lambda _{{\rm L}}\right] =\left[ 0.733\text{ 
}{\rm GeV},-0.6631\text{ }{\rm GeV},1~{\rm GeV}\right] ,  \label{43}
\end{equation}
and quark masses

\begin{equation}
\left[ m_{c},m_{b}\right] =\left[ 1.500~{\rm GeV},4.905~{\rm GeV}\right] .
\label{44}
\end{equation}
The potential forms in (39), and (42) were also used by Kiselev in [15].
Further, they were also been used for $\psi $ and $\Upsilon $ data probing $%
0.1~{\rm fm}<r<1$~${\rm fm}$\ region [13].

Further, Motyka and Zalewski [29] used a nonnrelativistic potential with $%
\alpha =1$ and $\beta =\frac{1}{2}$ for the $\overline{b}b$ quarkonium and
then applied it to the $\overline{c}c$ and $\overline{b}c$ quarkonium
systems. Grant, Rosner, and Rynes [31] have suggested $\alpha =0.045,$ $%
\beta =0.$ Heikkila, T\"{o}rnquist and Ono [32] tried $\alpha =1,$ $\beta =%
\frac{2}{3}.$ Some very successful potentials known from the literature are
not of this type. Examples are the Indiana potential [33] and the Richardson
potential [34].

\subsection{QCD-motivated potentials\ \ \ \ \ \ \ \ \ \ \ \ \ \ \ \ \ \ \ \
\ \ \ \ \ \ \ \ \ \ \ \ \ \ \ \ \ \ \ \ \ \ \ \ \ \ \ \ \ \ \ \ \ \ \ \ \ \
\ \ \ \ \ }

We use two types of the QCD-motivated potentials: The Igi-Ono (IO) and an
improved Chen-Kuang (CK) potential models. The details of these potentials
can be tracesd in our previous works in [17].

\section{CALCULATIONS AND CONCLUSION}

Based on our previous works [17], we determine the position of the
charmonium center-of-gravity (c.o.g.) $\overline{M}_{\psi }({\rm 1S})$ mass
spectrum and its hyperfine splittings by fixing the coupling constant $%
\alpha _{s}(m_{c})$ in (17) for each central potential$.$ Further, we
calculate the corresponding low-lying (c.o.g.) $\overline{M}_{\Upsilon }(%
{\rm 1S})$ and consequently the low-lying $\overline{M}_{B_{c}}({\rm 1S}).$%
\footnote{%
Kiselev {\it et al.} [15] have taken into account that $\Delta M_{\Upsilon }(%
{\rm 1S})=\frac{\alpha _{s}(\Upsilon )}{\alpha _{s}(\psi )}\Delta M_{\psi }(%
{\rm 1S})$ with $\alpha _{s}(\Upsilon )/\alpha _{s}(\psi )\simeq 3/4.$ On
the other hand, Motyka and Zalewiski [29] also found $\frac{\alpha
_{s}(m_{b}^{2})}{\alpha _{s}(m_{c}^{2})}\simeq 11/18.$} In Table I, we
estimate the radial wave function of the low-lying state of the $\overline{b}%
c$ system, so that

\begin{equation}
\left| R_{{\rm 1S}}(0)\right| =1.18-1.24~{\rm GeV}^{{\rm 3/2}},  \label{45}
\end{equation}
for the set of the central potentials given in section IVA. Further, in
Table II we present our results for the NR leptonic constant $f_{B_{c}}^{%
{\rm NR}}=466\pm 19$ ${\rm MeV}$ and $\ f_{B_{c}^{\ast }}^{{\rm NR}}=464\pm
19$ ${\rm MeV}$ as an estimation of the potential models without the
matching. Our results are compared with those of Gershtein, Likhoded and
Sabespi [35], who used Martin's potential and with those of Jones and
Woloshyn [36]. Moreover, the one-loop correction\ $\ f_{B_{c}}^{{\rm (1-loop)%
}}$ and the two-loop correction $\ f_{B_{c}}^{{\rm (2-loop)}}$ are also
given in Table III. Therefore, in the view of our results, our prediction
for the one-loop calculation is

\begin{equation}
\ f_{B_{c}}^{{\rm (1-loop)}}=408\pm 16~{\rm MeV,}  \label{46}
\end{equation}
and also for two-loop calculation

\begin{equation}
\ f_{B_{c}}^{{\rm (2-loop)}}=315\pm 50~{\rm MeV.}  \label{47}
\end{equation}
Our estimation for \ $f_{B_{c}}^{{\rm NR}}$ is fairly in good agreement with
the estimates in the framework of the lattice QCD result [8], $\ f_{B_{c}}^{%
{\rm NR}}=440\pm 20$ ${\rm MeV},$ QCD sum rules [9], potential models [2]
and also the scaling relation [15]. It indicates that the one-loop matching
in Ref. [26] provides the magnitude of correction of nearly $12\%.$ However,
the recent calculation in the heavy quark potential in the static limit of
QCD [13] with the one-loop matching [14] is

\begin{equation}
f_{B_{c}}^{{\rm (1-loop)}}=400\pm 15~{\rm MeV},  \label{48}
\end{equation}
Therefore, in contrast to the discussion given in [14], we see that the
difference is not crucially large in our estimation to one-loop value in the 
$B_{c}$-meson.

Our final result for the two-loop calculations is

\begin{equation}
f_{B_{c}}^{{\rm (2-loop)}}=315_{-50}^{+26}~{\rm MeV},  \label{49}
\end{equation}
the larger error value in (49) is due to the strongest running coupling
constant in Cornell potential.

Slightly different additive constants is permitted in this sector for a
charmed mesons to bring up data to its (c.o.g.) value. However, with no
additive constant to the Cornell potential [37], we notice that the smaller
mass values for the composing quarks of the meson leads to a rise in the
values of the potential parameters which in turn produces a notable lower
value for the leptonic constant as seen in Tables II and III.

On the other hand, for the QCD motivated two types IO potential, our
calculations for the ground state radial wave functions are presented in
Table IV. In Table V, the NR leptonic constant of the pseudoscalar $B_{c}$
state are in the range $354-426$ ${\rm MeV}$, and for the\ $B_{c}^{\ast }$
are $353-424$ ${\rm MeV},$ for the type I. Further, they are found $353-457$
and $351-454,$ for the type II. For instance, we may choose $f_{B_{c}}^{{\rm %
NR}}=391$ ${\rm MeV}$ with $\Lambda _{\overline{MS}}=200$ ${\rm MeV},$ for
the type I, and $f_{B_{c}}^{{\rm NR}}=396$ ${\rm MeV}$ with $\Lambda _{%
\overline{MS}}=300$ ${\rm MeV},$ for the type II. These results surpass the
matching procedure as well as methods followed in other references, cf.
Refs. [8,13,14]. Further, for the CK potential, we present the decay
constant in Table VI.\ 

The scaling relation (SR) for the ${\rm S}$-wave heavy quarkonia has the
form [15]

\begin{equation}
\frac{f_{n}^{2}}{M_{n}(\overline{b}c)}\left( \frac{M_{n}(\overline{b}c)}{%
M_{1}(\overline{b}c)}\right) ^{2}\left( \frac{m_{c}+m_{b}}{4\mu }\right) =%
\frac{d}{n},  \label{50}
\end{equation}
where $m_{c}$ and $m_{b}$ are the masses of heavy quarks composing the $%
B_{c} $-meson, $\mu $ is the reduced mass of quarks, and$\ d$ is a constant
independent of both the quark flavors and the level number $n.$ The value of 
$\ d$ is determined by the splitting between the ${\rm 2S}$ and ${\rm 1S}$
levels or the average kinetic energy of heavy quarks, which is independent
of the quark flavors and $n$ with the accuracy accepted. The accuracy
depends on the heavy quark masses and it is discussed in [15] in detail. The
parameter value in (50), $d\simeq 55$ ${\rm MeV},$ can be extracted from the
experimentally known leptonic constants of $\psi $ and $\Upsilon $ [15].
Hence, in the view of Table VII, the SR gives for the ${\rm 1S}$-level

\begin{equation}
\ f_{B_{c}}^{{\rm (SR)}}\simeq 444_{-23}^{+6}~{\rm MeV},  \label{51}
\end{equation}
for all static potentials used. Kiselev estimated $\ f_{B_{c}}^{{}}=400\pm
45~{\rm MeV}$ [14] and \ $f_{B_{c}}^{{\rm (SR)}}=385\pm 25~{\rm MeV}$ [15],
Narison found \ $f_{B_{c}}^{{\rm (SR)}}=400\pm 25~{\rm MeV}$ [38].

Overmore, we give the leptonic constants for the excited ${\rm nS}$-levels
of the $\overline{b}c$ quarkonium system in Table VII. We remark that the
calculated value of $\ f_{B_{c}(2S)}^{{\rm (SR)}}=300\pm 15~{\rm MeV}$ is in
good agreement with $\ f_{B_{c}{\rm (2S)}}^{{\rm (SR)}}=280\pm 50~{\rm MeV}$
being calculate in [13] for the ${\rm 2S}$-level in the $\overline{b}c$
system and it is also consistent with the scaling relation [15].

In Figure 1, we plot the calculated values of $B_{c}$ leptonic constants
using (50) for different potential models together with the calculated
values of the excited ${\rm nS}$-states using [15]

\begin{equation}
f_{n}=\frac{1}{\sqrt{n}}f_{1}.  \label{52}
\end{equation}
Hence, our findings obtained from the potential model and SR are in good
agreement to several ${\rm MeV}$ as shown in Figure 1.

Finally, we have also noted that the leptonic constant is practically
independent of the total spin of quarks, so that

\begin{equation}
f_{V,n}\simeq f_{P,n}=f_{n},  \label{53}
\end{equation}
where $M_{B_{c}}\simeq m_{b}+m_{c}.$ Thus, one can conclude that for the
heavy quarkonium, the QCD sum rule approximation gives the identity of
leptonic constant values for the pseudoscalar and vector states.

In this work, we have successfuly applied the SLNET using a class of static
and QCD-motivated potentials to calculate numerically the leptonic constant
of the pseudoscalar and vector $B_{c}$-meson. Once the experimental leptonic
constant of the $B_{c}$ meson becomes clear, one can sharpen the analysis.

\acknowledgments
The author is grateful to Prof. Dr. \c{S}enol Bekta\c{s} the vice president
of the NEU for the encouragement to do the work, to Prof. Ramazan Sever in
METU for the enlightening discussions and also Dr. Suat G\"{u}nsel the
founder president of the NEU for the financial support and assistance.

\bigskip

\bigskip \baselineskip= 2\baselineskip

\mediumtext

\begin{table}[tbp]
\caption{The characteristics of the wave function $\left| \protect\psi _{%
{\rm 1S}}(0)\right| ^{2}$ and the radial wave function $\left| R_{{\rm 1S}%
}(0)\right| ^{2}=4\protect\pi \left| \protect\psi _{{\rm 1S}}(0)\right| ^{2}$
both at the origin (in ${\rm GeV}^{{\rm 3}})$ obtained from the
Schr\"{o}dinger equation for various potentials.}
\label{table 1}
\begin{tabular}{lllllll}
Level & Cornell & Song-Lin & Turin & Martin & Logarithmic & Cornell%
\tablenotemark[1] \\ 
\tableline$\left| \psi _{1S}(0)\right| ^{2}$ & 0.112 & 0.123 & 0.111 & 0.119
& 0.102 & 0.065 \\ 
$\left| R_{1S}(0)\right| ^{2}$ & 1.413 & 1.54 & 1.397 & 1.495 & 1.28 & 0.814
\end{tabular}
\tablenotetext[1]{Here we cite Ref. [37] for the fitted set of parameters.}
\end{table}

\mediumtext

\bigskip 
\begin{table}[tbp]
\caption{Pseudoscalar and vector decay constants $%
(f_{P}=f_{B_{c}},~f_{V}=f_{B_{c}^{\ast }})$ of the $B_{c}$ meson$,$
calculated in the different potential models (the accuracy is 5$\%$), in $%
{\rm MeV}$. }
\label{table2}
\begin{tabular}{lllllllll}
Constants & Cornell\tablenote{Here $V_{0}\ne0$.} & Song-Lin & Turin & Martin
& Logarithmic & Cornell\tablenote{Here $V_{0}=0$, see Ref. [37].} & [35] & 
[36] \\ 
\tableline$f_{B_{c}}^{NR}$ & 464.5 & 485.1 & 462.0 & 478.1 & 441.7 & 351.5 & 
460$\pm 60$ & 420$\pm 13$ \\ 
$f_{B_{c}^{\ast }}^{NR}$ & 461.5 & 482.2 & 459.2 & 475.3 & 439.3 & 349.7 & 
460$\pm 60$ & 
\end{tabular}
\end{table}

\bigskip

\bigskip

\mediumtext

\bigskip 
\begin{table}[tbp]
\caption{One-loop and two-loop corrections to pseudoscalar and vector decay
constants of the low-lying $\overline{b}c$ system$,$ calculated in the
different potential models (the accuracy is 4$\%$), in ${\rm MeV}$. }
\label{table3}
\begin{tabular}{lllllll}
Quantity & Cornell & Song-Lin & Turin & Martin & Logarithmic & Cornell \\ 
\tableline$f_{B_{c}}^{(1-loop)}$ & 393.6\tablenote{To the first correction
for all potentials, $K_{0}=0.85-0.90$.} & 424.4 & 399.6 & 421.2 & 399.3 & 
311.9 \\ 
$f_{B_{c}}^{(2-loop)}$ & 264.1\tablenote{To the second correction for all
potentials, $K_{0}=0.57-0.77$.} & 333.0 & 296.6 & 339.2 & 340.9 & 238.1 \\ 
$f_{B_{c}^{\ast }}^{(1-loop)}$ & 391.0 & 421.9 & 397.1 & 418.7 & 397.1 & 
310.3 \\ 
$f_{B_{c}^{\ast }}^{(2-loop)}$ & 262.4 & 331.0 & 294.8 & 337.2 & 339.0 & 
236.9
\end{tabular}
\end{table}

\bigskip \bigskip \bigskip \bigskip

\mediumtext

\bigskip 
\begin{table}[tbp]
\caption{The characteristics of the radial wave function at the origin $%
\left| R_{{\rm 1S}}(0)\right| ^{2}$ (in ${\rm GeV}^{{\rm 3}})$ obtained from
the Schr\"{o}dinger equation for the Igi-Ono potential model.}
\label{table 4}
\begin{tabular}{llllll}
&  &  & $\Lambda _{\overline{MS}}$ &  &  \\ 
Level\tablenotemark[1] & 100 & 200 & 300 & 400 & 500 \\ 
\tableline Type I\tablenotemark[2] &  &  &  &  &  \\ 
$\left| \psi _{1S}(0)\right| ^{2}$ & 0.066 & 0.08 & 0.092 & 0.095 & 0.089 \\ 
$\left| R_{1S}(0)\right| ^{2}$ & 0.826 & 1.005 & 1.156 & 1.19 & 1.114 \\ 
\tableline Type II\tablenotemark[3] &  &  &  &  &  \\ 
$\left| \psi _{1S}(0)\right| ^{2}$ & 0.065 & 0.071 & 0.082 & 0.096 & 0.109
\\ 
$\left| R_{1S}(0)\right| ^{2}$ & 0.819 & 0.891 & 1.03 & 1.204 & 1.366
\end{tabular}
\tablenotetext[1]{For the fitted parameters, see Ref. [17].}%
\tablenotetext[2]{We shifted the $\bar{c}c$ spectra by $c=-22$ to $-31$
$MeV$.}%
\tablenotetext[3]{We shifted the $\bar{c}c$ spectra by $c=-15$ to $-26$
$MeV$.}
\end{table}

\mediumtext

\bigskip

\bigskip 
\begin{table}[tbp]
\caption{Pseudoscalar and vector decay constants of the $B_{c}$ meson$,$
calculated in the Igi-Ono potential model, in ${\rm MeV}$. }
\label{table 5}
\begin{tabular}{llllll}
&  &  & $\Lambda _{\overline{MS}}$ &  &  \\ 
Quantity\tablenotemark[1] & 100 & 200 & 300 & 400 & 500 \\ 
\tableline Type I\tablenotemark[2]:$\alpha _{s}=$ & 0.199\tablenotemark[3] & 
0.217 & 0.238 & 0.25 & 0.262 \\ 
$f_{B_{c}}^{NR}$ & 354.1 & 391.1 & 420.0 & 426.0 & 411.7 \\ 
$f_{B_{c}^{\ast }}^{NR}$ & 352.6 & 389.2 & 417.7 & 423.6 & 409.4 \\ 
\tableline Type II\tablenotemark[4]:$\alpha _{s}=$ & 0.199 & 0.227 & 0.230 & 
0.241 & 0.251 \\ 
$f_{B_{c}}^{NR}$ & 352.7 & 368.2 & 396.1 & 428.6 & 456.9 \\ 
$f_{B_{c}^{\ast }}^{NR}$ & 351.2 & 366.4 & 394.1 & 426.4 & 454.4
\end{tabular}
\tablenotetext[1]{For the fitted parameters we cite Ref. [17].}%
\tablenotetext[2]{We shifted the $\bar{c}c$ spectrum by $c=-22$ to $-31$
$MeV$.}%
\tablenotetext[3]{Input value
$\Delta E_{HF}$$=117$ $MeV$; determines $\alpha_{s}(m_{c})$.}%
\tablenotetext[4]{We shifted the $\bar{c}c$ spectra by $c=-15$ to $-26$
$MeV$.}
\end{table}

\bigskip \mediumtext

\bigskip

\bigskip 
\begin{table}[tbp]
\caption{The radial wave function, pseudoscalar and vector decay constants
of the $B_{c}$ meson$,$ calculated in the improved Chen-Kuang potential
model, in ${\rm MeV}$. }
\label{table 6}
\begin{tabular}{lllllllll}
&  &  &  & $\Lambda _{\overline{MS}}$ &  &  &  &  \\ 
Quantity\tablenotemark[1] & 100 & 180 & 300 & 350 & 375 & 400 & 450 & 500 \\ 
\tableline$\left| R_{1S}(0)\right| ^{2}$ & 1.017 & 1.017 & 1.017 & 1.017 & 
1.017 & 1.015 & 0.957 & 0.829 \\ 
$f_{B_{c}}^{NR}$ & 393.5 & 393.5 & 393.5 & 393.4 & 392.5 & 389.9 & 381.8 & 
353.8 \\ 
$f_{B_{c}^{\ast }}^{NR}$ & 391.4 & 391.4 & 391.4 & 391.3 & 390.4 & 387.8 & 
379.8 & 352.3
\end{tabular}
\tablenotetext[1]{Input value
$\Delta E_{HF}$$=117$ $MeV$; determines $\alpha_{s}(m_{c})=0.27$.}
\end{table}

\bigskip

\mediumtext

\bigskip 
\begin{table}[tbp]
\caption{The ${\rm nS}$-levels leptonic constant of the $\overline{b}c$
system$,$ calculated in the different potential models (the accuracy is $%
3-7\%$), in ${\rm MeV}$, using the SR. }
\label{table7}
\begin{tabular}{llllll}
Quantity & Cornell & Song-Lin & Turin & Martin & Logarithmic \\ 
\tableline$f_{1S}$ & 449.6 & 450.4 & 448.0 & 448.8 & 420.9 \\ 
$f_{2S}$ & 305.8 & 305.0 & 303.3 & 303.5 & 284.7 \\ 
$f_{3S}$ & 243.0 & 243.2 & 241.3 & 241.8 & 227.2 \\ 
$f_{4S}$ & 206.0 & 207.1 & 204.9 & 205.9 & 193.8
\end{tabular}
\end{table}
\bigskip

\bigskip

\begin{figure}[tbp]
\caption{The $n{\rm S}$-levels leptonic constant of the $B_{c}$ system
calculated in the different potential models using SR.}
\label{autonum}
\end{figure}
\begin{verbatim}
 
\end{verbatim}

\end{document}